\documentstyle[12pt,epsf]{article}
\begin{document}
\vglue 2truecm
\begin{center}
{\Large
\bf  Energy level statistics of electrons in a
 2D quasicrystal \\}

\vspace*{1.5cm}
{\bf Fr\'ed\'eric Pi\'echon and Anuradha Jagannathan\\

\vspace*{1cm}
Laboratoire de Physique des
Solides, associ\'e au CNRS \\
Universit\'{e} Paris--Sud \\ 91405 Orsay, France \\
}
\vspace*{1.4cm}
\vspace*{1.4cm}
\end{center}

A numerical study is made of the spectra of a tight-binding
hamiltonian on square approximants of the quasiperiodic octagonal tiling.
Tilings may be pure or random, with different degrees of phason disorder
considered.
The level statistics for the randomized tilings follow the predictions
of random
matrix theory, while for the perfect tilings a new type of level
statistics is found. In this case,
 the first-, second-
level spacing distributions are well described
by lognormal laws with power law tails for large spacing. In addition,
level spacing properties being related to properties of the density of
states, the latter quantity is studied and the multifractal character of
the spectral measure is exhibited.
\vskip 3cm



\vfill
\eject

Since the discovery of quasicrystalline alloys, many geometrical
models have been built that are consistent
with observed diffraction patterns. Tilings may
have perfect quasiperiodic long range order with local symmetries that
are not allowed for ordinary periodic structures, and which give rise to
the five-fold, eight-fold or ten-fold symmetries of the diffraction
patterns. Alternatively, one may consider
randomized versions of perfect tilings.
The approximants can be geometrically disordered to varying degrees
by performing
``phason'' flips whereby a vertex is displaced and the tiling redrawn,
resulting in a rearrangement of the tiles that surround that vertex.
``allowed" permutations of the tiles. A typical random tiling
(i.e. excluding some special cases that could be generated by the
randomization procedure) has the same local symmetry as its perfect
counterpart. The perfect tiling of course has exact
symmetries such as its
self-similarity property under inflation or deflation operations, while
random tilings do not have this property.
Their orientational order and the constraint imposed by the
fixed shape of the tiles of the random tiling are sufficient, however,
to give rise to sharp peaks in its diffraction pattern, similar to
the Bragg peaks of the perfect tiling \cite{hen}.
Many existing phenomenological
models to explain quasicrystal electronic properties are
based on the construction of a pseudo Brillouin zone taking into account the
brightest spots of the diffraction pattern, and at this level they
would thus not distinguish
between the perfect and random tiling models.
The effect of phason disorder on electronic properties can however be very
profound, as will be shown below.

In this paper we make a numerical study of statistical properties of the
eigenvalue spectrum of a tight-binding Hamiltonian describing the hopping of
electrons on vertices of a two dimensional tiling. Hopping is allowed
between sites that are separated by one unit length (the tile side is taken
to be of length one). The tilings are chosen to be square periodic
approximants of the quasiperiodic octagonal
tiling, containing upto 8119 sites (the largest approximant considered
here).
The hopping parameter is taken to be constant and set equal to
unity, so that the resulting spectra are a pure consequence of
the geometry, or site connectivity, of the tilings. The
Hamiltonian is diagonalized numerically for each case, and the energy
levels are used to compute the distribution $P(s)$ of spacings of
 $n^{th}$ nearest-neighbor energy levels $s=E_{i+n}-E_{i}$ (supposing
the levels to be ordered). To place the current study in perspective,
we mention previous work on
electronic levels of a tight-binding model, namely studies of
disordered $periodic$ lattices on both sides of the metal-insulator
transition\cite{mon}. These show that in the metallic regime,
the first-neighbor level spacing distributions P(s)
have the forms of the Wigner distributions, obtained
for gaussian random matrices\cite{bopm}. In other words
the energy levels of a real symmetric Hamiltonian with random
distributed onsite energies have the statistics of levels of the gaussian
orthogonal ensemble (GOE), where $P(s) \sim
s^\beta \exp (-c_\beta s^2)$ where $c_\beta$ is a known constant
 and $\beta=1$. In the case of a magnetic flux traversing
the lattice,
the Hamiltonian belongs in the unitary class and levels obey gaussian
unitary ensemble (GUE) statistics, with $\beta=2$. Finally, in the
insulating regime (strong disorder)
P(s) has a Poissonian form for large spacings, $P(s) \sim \exp (-s)$.
A second quantity of
interest calculated for these crystalline models
is the spectral rigidity $\Sigma^{2} (E)$, which measures
the fluctuation of the number of levels in a energy window of width $E$.
$\Sigma^{2} (E)$ grows logarithmically
with $E$ at small $E$ in the metallic diffusive
regime \cite{mon}, just as for the random matrix ensembles.
This behavior is a consequence of correlations between
levels, and gives a smaller
$\Sigma^{2} (E)$
compared to that of uncorrelated levels (as occurs in the insulating
regime).

In our present work, we find that the randomized quasiperiodic tiling has
level statistics of the GOE or GUE type, depending on the applied flux.
  The statistics of energy levels are thus
 similar to those of disordered tight binding models on
crystals in the diffusive regime, even though in our case the disorder is
purely geometric \cite{note}.

For the perfect approximants, we find a
new type of level statistics. The perfect octagonal approximant has
been studied previously by Benza and Sire \cite{ben} who presented
results on level statistics. They did not propose a fit to any functional
form, noting principally that
level repulsion occurs in the perfect quasicrystal, since P(s) drops to
zero at small values of s. This is certainly true of the distribution
that we propose in this paper. For the perfect case, the calculations show
that the density of states (DOS) has
huge fluctuations at all energy scales, indicating
multifractal
properties of the spectral measure. We have calculated the effective
$f(\alpha)$ function of
local singularity exponents $\alpha$.
Its maximum occurs at
 $\alpha= D_I = f(D_I)$ (the most probable dimension) and
has the maximum value $f_{max}= D_{F} \sim 1$ (the fractal
dimension). The last is an indication that the spectrum is gapless, in
accord with the finding of \cite{ben}. The calculated first-neighbor
distribution  $P(s)$, as well as the second- and third-neighbor
spacing distributions $P^{(2)}(s)$ and $P^{(3)}(s)$
are well described by lognormal
laws with power law tails at large spacings. To complicate matters,
however, the low moments of the spacing distributions apparently
do not have the size dependence that would be indicated by the
calculated $f(\alpha)$, while the high moments have a size dependence
that is characteristic of a power law behavior $s^{-\gamma}$ in the
large $s$ tail of P(s). It should be noted that the systems
studied are comparatively small, and studying the next size of
approximant will
help in determining better the size dependence of these distributions.

We now describe more precisely our
calculations. The perfect periodic approximant is
generated using the method described by Duneau $et \ al$ \cite{dom}. The
largest of the three sizes of tiling used in these computations
is the 5/4 approximant which contains 8119 atoms in the unit cell.
The random
phason disordered approximants are generated by an iterative process of
displacing sites and computing the new links between sites in each
cycle. The
Hamiltonian for a given tiling is defined by the following site-projected
form
\begin{equation}
{ (H\psi)_{i}=\sum_{<ij>} \psi_{j}=E\psi_{i} }
\label{e2}
\end{equation}
\noindent
where the sum is taken over sites j linked to site i. The number of links
emanating from a given
site i varies between 3 and 8 in the perfect case, corresponding to the
six local environments \cite{dun} of the octagonal tiling.
In the randomized case new local configurations appear, although
 the mean coordination is $\overline{z}=4$ in both tilings. To reiterate, in
this model the quasiperiodicity as well as the disorder are
present purely in the connectivity and are of purely geometric nature. This
may be compared with models in which
one takes an explicit parameter in
the Hamiltonian to generate the quasiperiodicity or disorder, such as a
site-dependent potential. This distinction is not possible in one dimension,
where the quasi-periodic nature of connections between sites must be coded
by introducing for example two kinds of bonds, as in the Fibonacci chain
\cite{stein}.
Periodic continuation of the $L \times L$  square tilings allows to
impose boundary conditions of the form
\(\psi(x+L,y)=e^{i\phi}\psi(x,y)\); $\psi(x,y+L)=\psi(x,y)$. This is
equivalent to adding a
 magnetic flux $\phi$ along the $y$ or $z$ axis. It is thus possible to
eventually induce a GOE to GUE
transition by varying $\phi$.

 We discuss first the results for the geometrically disordered case.
Their analysis and interpretation is more straightforward and so will
serve as a guide to the case of the perfect approximant.
 The numerical results obtained
for $P(s)$ (taking a normalized spacing $s= (E_{i+1}-E_i)/(W/N)$ where W is
the band width and N the number of levels) and for $\Sigma^{2}(E)$
are shown in Figs. 1a and 1b. The points are obtained for flux values
$\phi=0$ and $\phi=1/2$ , while the curves give
the fit to the corresponding analytical expressions from
random matrix theory (RMT). It has been shown, for weakly disordered metals
that the RMT law for the spectral stiffness
can be derived, both in perturbation theory \cite{altsk}, and
by means of
 a semiclassical model \cite{imsha} by $assuming$ a diffusive motion
for the particle. Both derivations obtain that the spectral stiffness
follows the RMT logarithmic law in the low energy region
$\Sigma^{(2)}(E)
\propto \frac{2}{\beta \pi^{2}}\ln(E) + cst $, crossing over to a linear
behavior (in two dimensions). Our calculations yield a similar behavior
of the spectral rigidity. This may indicate that the
disordered  quasicrystal possesses a diffusive regime, as for the disordered
crystalline metals. More precisely, if the wavefunctions are
localized in the quasiperiodic tiling,
that the localisation length must be greater than the system size, since
on the scale of our samples the results are in good accord with the Wigner
distributions.
We note that we have not been able to disorder (in the geometric sense) the
tilings sufficiently
to see a crossover to strong localization. Such a crossover can be seen in
tilings that have a strong $energetic$ disorder (see note in \cite{note})
wherein we take
a onsite energy randomly from an energy interval of width comparable to the
bandwidth
of the perfect tiling. In the strongly localized r\'egime, $P(s)$ tends to
the exponentially decaying Poissonian form \cite{bopm}.
Our results indicate that
in all likelihood, the geometrically randomized tilings still possess
sufficient
long range order that wavefunctions remain ``extended" in the sense of not
decaying exponentially, however, this remains to be established.

In contrast to the random case where the fluctuations of the
DOS are small compared to Poissonian, the perfect approximant,
like one dimensional quasiperiodic systems such as the Fibonacci chain,
has a self-similar DOS with huge fluctuations at all energy scales
\cite{tsu}.
Quantities which characterize these multifractal properties are the
generalized fractal dimensions $D(q)$. To find these we consider
 a partition of the bandwidth $W$ into $M$
distinct boxes $S_{i}$ of probability $p_{i}$ and size $l_{i}$ where
$p_{i}=\frac{n_{i}}{N}$ and $n_{i}$ is the number of level in the box of
size $l_{i}$. The dimensions $D(q)$ satisfy the following condition
(\cite{hal})
\begin{equation}
\sum_{i=1}^{M} \frac{ p_{i}^{q-1} }{ l_{i}^{\tau} } \approx 1
\label{b4}
\noindent
\end{equation}
with $\rm {max}_{i}(l_{i}) \rightarrow 0$
and $\tau(q)=(q-1)D(q)$.
In the simplest case, one takes a partition of boxes with
equal size $l=W/M$. In that
case, the
measure $p_{i}$ is assumed to scale as $p_{i} \sim l^{\alpha_{i}}$
when $l$ tends to zero and
the number of $\alpha_{i}$ between $\alpha$ and $\alpha+d\alpha$ is
assumed to vary as $N_{l}(\alpha) \sim l^{-f(\alpha)}d\alpha$.
The function $\tau(q)$ vs $q$  and the effective
$f(\alpha)$ vs $\alpha$ are related by means of a Legendre transform.
Conversely, one can fix the weights of the boxes to be identical,
allowing their widths to fluctuate, and calculate $q(\tau)$. This is
tantamount to calculating moments of spacings with respect to the
underlying distributions $P^{(n)}(s)$.
The $n^{th}$ neighbor level spacing moments
$
l_{i}=
{s_{i}^n=(E_{i+n}-E_{i})/W,i=1,N}$ are obtained by taking the partition
of fixed probability $p=n/N$.

We have computed the
$f(\alpha)$ function
using the two methods described above.
 In agreement with the earlier work of Benza and Sire
who found a single-band spectrum for
the pure hopping model, we find the
maximum of $f(\alpha)$ to have the value
$D_F \sim 1$ indicating the spectrum is gapless. Another numerical
argument
in favor of a spectrum with no (finite measure) gap is our finding that
the biggest spacing $s_{max}$ scales as $\sim N^{-1/2}$
($s_{max} \rightarrow 0$ for increasing $N$ is a necessary
condition
to apply the second of the two methods described above).
The information dimension defined earlier is $D_I=0.98$
and a fit to a parabolic form of $f(\alpha)$ close to its maximum gives
its curvature $1/\mu$,
$\mu=0.03$. Now a parabolic shape for $f(\alpha)$ can be shown to lead to
lognormal spacing distributions for each of the $nth$ nearest neighbor
spacings $s^n$ (for $n << N$). The parameters of these distributions are
related
to $D_I$ and $\mu$. The distributions depend, as well, upon the system
size.  We have found, however, that when the distribution $P(s)$ of
rescaled variables $s=N(E_{i+1}-E_{i}/W)$ is plotted, a single
curve is obtained for the three sizes studied: $N=$239,1393 and 8119 sites.
The same remark applies to the two other distributions calculated,
$P^{(2,3)}$. A comment on the calculation
of $P(s)$: it is well known that before computing any statistical spectral
properties, the spectrum has to be unfolded \cite{bopm}.
This procedure can be understood within a semiclassical scheme
 as being equivalent to taking only
the fluctuating part of the integrated density of states. One keeps only the
quantum
interference corrections around the Thomas-Fermi zeroth $\hbar$ order
term of the integrated density of states.
However to have an efficient unfolding, the fluctuating part has to be
a small perturbation around the zeroth $\hbar$ order term and the zero
order term itself should not be strongly fluctuating. In weakly disordered
crystalline systems this is the case.
In contrast, the density of state of a quasiperiodic system
like ours has huge fluctuations at all energy scales. One may consider that
in this case at least one of the hypothesis breaks.
Thus the unfolding procedure
is not well defined for our spectrum. We have therefore computed $P(s)$
without any unfolding.

The distribution of the logarithm
of $s$ is well fitted by a gaussian, so that $P(s)$ is a
lognormal function of the form:
\begin{equation}
P(s) =  \frac{1}{\sqrt{\pi B} s}exp-\frac{(ln(s)-ln(s_{0}))^{2}}{2B}
\label{f1}
\end{equation}
with $ln(s_{0})=-B/2$ for variables normalized to unity so that
$\int_{0}^{\infty}s P(s)ds=1$. Level repulsion clearly occurs,
as $P(s)$ tends to zero with diminishing $s$ at small $s$. The lognormal
form is obeyed for spacings around the most probable value.At small $s$
the repulsion seems $linear$ rather than $exponential$ as given by
the lognormal, while at large spacings, $s\geq3W/N$, the error function
$\int_{s}^{\infty}P(s)ds$ behaves as $\sim s^{-2}$ implying $P(s)\sim
s^{-3}$ (this agrees with the observed variation in the largest
spacing as a function of the system size and with $P(s) \sim
s^{-(1+\alpha_{max})}$). A final observation on the dependence of $P(s)$ on
the flux traversing the tiling: we have done a separate calculation for each
value of the flux $\phi$. In the perfect approximants we find $P(s)$
to be independent
of the flux, unlike the case of weakly localized systems
\cite{mon}.

In conclusion, we have numerical evidence for fundamental differences
underlying spectra of perfect and random quasicrystal approximants, the
latter resembling disordered crystalline systems in so far as their
level statistics are concerned. The perfect approximants have lognormal
distributions for the level spacings, with crossover to a power law in
the tails. Results for different sample sizes can be made to coincide
by plotting distributions of suitably scaled variables. This result
seems inconsistent with the size-dependence expected of a straight-forward
multifractal energy spectrum and calculations on bigger systems should
help resolve the issue.\\
{\it Acknowledgments}: We thank Gilles
Montambaux for many useful
discussions and for the use of several numerical routines for the analysis
of energy levels. Pascale Launois and Manfred Fettweis kindly checked the
similarity of diffraction patterns corresponding to perfect and randomized
tilings.
\\ 
\newpage
{\bf Figure Captions }\\
1a. Level spacing distribution for the randomized approximants; in the
GOE case ($\phi=0$ circles) and GUE ($\phi=1/2$ stars).
Continuous
lines show the corresponding Wigner distributions.\\
1b. Spectral rigidity in the same two cases along with the RMT fit (
filled circles: GOE, open circles: GUE). \\
2. $f(\alpha)$ distribution of singularity exponents $\alpha$ along with a
fit to a parabolic shape.\\
3a. Level spacing distribution for the perfect approximant fit to a
lognormal form. In the inset the distribution of the logarithm of $s$ is fit
to a gaussian.\\
3b. and 3c. Same as 3a for the next nearest and third neighbor spacing
distribution.\\

\begin{thebibliography}{10}
\bibitem{note} We have
also checked the effect of an energy disorder, by adding to the perfect
approximant an onsite
energy generated by a random numbers program, and verified that in this case
too, the Wigner forms are obtained for $P(s)$.
\end{thebibliography}
\end{document}